\documentclass[12pt, prd, showpacs]{revtex4}
\input{tcilatex}

\begin{document}

\title{Exactly solvable model of wormhole supported by phantom energy}
\author{O. B. Zaslavskii}
\affiliation{Department of Mechanics and Mathematics, Kharkov V.N.Karazin National
University, \\
Svoboda Square 4, Kharkov 61077, Ukraine}
\email{ozaslav@kharkov.ua}

\begin{abstract}
We have found a simple exact solution of spherically-symmetrical Einstein
equations describing a wormhole for an inhomogeneous distribution of the
phantom energy. The equation of state is linear but highly anisotropic:
while the radial pressure is negative, the transversal one is positive. At
infinity the spacetime is not asymptotically flat and possesses on each side
of the bridge a regular cosmological Killing horizon with an infinite area,
impenetrable for any particles. This horizon does not arise if the wormhole
region is glued to the Schwarzschild region. In doing so, the wormhole can
enclose an arbitrary amount of the phantom energy. The configuration under
discussion has a limit in which the phantom energy turns into the string
dust, the areal radius tends to the constant. In this limit, the strong
gravitational mass defect is realized in that the gravitational active mass
is finite and constant while the proper mass integrated over the total
manifold is infinite.
\end{abstract}

\maketitle

% It is always \today, today, but any date may be explicitly specified


Wormholes as spacetimes connecting two different universes via a throat and
having no horizons were invented pure theoretically \cite{mt} (see also a
textbook \cite{vis} and references therein). Nowadays an interest to such
objects has arisen significantly in connection with discovery of
acceleration of the Universe \cite{reiss}, \cite{per}. since in both cases
the usual energy conditions are violated. Thus, an interesting and
unexpected overlap between two seemingly different subjects appeared. More
precisely, wormholes are supported by so-called exotic matter with a
negative pressure $p<0$, and violation of a null energy condition, so that $%
p+\rho <0$ ($\rho $ is the energy density). The acceleration of the Universe
is caused by a hypothetical dark energy with the energy density $\rho >0$
and $p=-\kappa \rho $ with $k>\frac{1}{3}$. The important case of the dark
energy is the phantom energy \cite{cal} when $\kappa >1$ and the same
violation of the null energy condition as in the wormhole case occurs.
Meanwhile, the essential difference between two issues under discussion is
that in cosmology $p$ and $\rho $ depend usually on time but not on a
spatial coordinate, whereas for static wormhole the situation is opposite.
Although a variety of wormhole solutions were found for ghost scalar fields
starting from early works \cite{sc} (the recent situation is described, for
example, in \cite{shgr}), analysis for the phantom energy in terms of the
explicit equation of state has been carried out only recently \cite{sh}
(properties of wormholes supported by phantom energy were also studied in 
\cite{lobo}). Two models (with a constant energy density and a Gaussian
distribution) were analyzed and some restrictions on the size of the region
filled with the phantom energy or parameters of the model have been derived.
Such an approach enables one to look at subtle details of the wormhole
metric and spatial distribution of the phantom energy but, unfortunately, it
requires the function $\rho (r)$ to be defined "by hand" and no obvious
sensible relation between a transversal pressure and the energy density was
obtained. It looks more physically reasonable to define beforehand an
equation of state and find the solutions afterwards.

Such a programme is realized in our communication. Starting with a linear
equation of state, we demonstrate explicit exact solutions describing a
wormhole. In doing so, our equation of state is anisotropic in that the
transversal pressure $p_{\perp }$ does not coincide with the radial one $%
p_{r}$.

The metric of the system reads

\begin{equation}
ds^{2}=-dt^{2}U+\frac{dr^{2}}{V}+r^{2}d\omega ^{2}\text{, }d\omega
^{2}=d\theta ^{2}+\sin ^{2}\theta d\phi ^{2}\text{, }V=1-b(r)/r\text{.}
\label{met}
\end{equation}%
Here the coordinate $r$ runs in the range $r_{0}\leq r<\infty $. To describe
a wormhole, the metric (\ref{met}) should obey some conditions \cite{mt}, 
\cite{vis}. The metric coefficient $U$ should be finite and non-vanishing in
the vicinity of $r_{0}$, the shape function $b(r)$ should satisfy the
relations%
\begin{equation}
b(r_{0})=r_{0}\text{,}  \label{br}
\end{equation}%
\begin{equation}
b^{\prime }(r_{0})<1\text{,}  \label{b'}
\end{equation}%
\begin{equation}
b(r)<r\text{, }r>r_{0}\text{.}  \label{b<}
\end{equation}%
It follows from 00 and 11 Einstein equations respectively that%
\begin{equation}
\frac{b^{\prime }}{8\pi r^{2}}=\rho (r)\text{,}  \label{00}
\end{equation}%
\begin{equation}
\frac{U^{\prime }}{U}=\frac{8\pi p_{r}r^{3}+b}{r(r-b)}\text{,}  \label{11}
\end{equation}%
where $T_{0}^{0}=-\rho $, $T_{r}^{r}=p_{r}$, $T_{\nu }^{\mu }$ is the
stress-energy tensor. Integration of eq. (\ref{00}) gives us the value of
the shape function%
\begin{equation}
b(r)=r_{0}+2m(r)\text{, }m(r)=4\pi \int_{r_{0}}^{r}d\tilde{r}\tilde{r}%
^{2}\rho (\tilde{r})\text{.}  \label{b}
\end{equation}

Then, by substitution of (\ref{b}) into (\ref{11}) one obtains that the
condition $U(r_{0})\neq 0$ entails 
\begin{equation}
p_{r}^{(0)}\equiv p_{r}(r_{0})=-\frac{1}{8\pi r_{0}^{2}}\text{.}  \label{p0}
\end{equation}%
It follows from (\ref{b}) and (\ref{b'}) that 
\begin{equation}
p_{r}^{(0)}+\rho _{0}<0\text{, }\rho _{0}\equiv \rho (r_{0})\text{.}
\label{p+}
\end{equation}

One can also derive from the conservation law of the stress-energy tensor $%
T_{\mu ;\nu }^{\nu }=0$ with $\mu =r$ that%
\begin{equation}
p_{\perp }=\frac{r}{2}[p_{r}^{\prime }+\frac{2p_{r}}{r}+\frac{U^{\prime }}{2U%
}(p_{r}+\rho )]\text{,}  \label{cl}
\end{equation}%
where $T_{\theta }^{\theta }=T_{\phi }^{\phi }=p_{\perp }$. From now on we
assume that our source is realized by the phantom energy with the equation
of state that contains a \textit{radial }pressure \cite{sh}

\begin{equation}
p_{r}=-\kappa \rho \text{, }\kappa >1\text{,}  \label{es}
\end{equation}%
where $p_{r}<0$. We suppose also that the pressures are anisotropic and%
\begin{equation}
p_{\perp }=\alpha \rho \text{.}
\end{equation}%
Thus, we consider a simplest case of a linear relation between pressure and
energy density but with $p_{\perp }\neq p_{r}$.

In general, even with specification made it is impossible to find exact
solutions of Einstein equations. However, it turns out that if we choose%
\begin{equation}
\alpha =\frac{\kappa -1}{4}>0\text{,}  \label{ak}
\end{equation}%
equations (\ref{00}), (\ref{11}) and (\ref{cl}) admit exact analytical
solutions%
\begin{equation}
\rho =\frac{d}{8\pi r^{2}}\text{, }d\equiv \frac{1}{\kappa }<1\text{.}
\end{equation}%
\begin{equation}
b=r_{0}+d(r-r_{0})\text{,}
\end{equation}%
\begin{equation}
ds^{2}=-dt^{2}\frac{r_{1}}{r}+\frac{dr^{2}}{(1-d)(1-\frac{r_{0}}{r})}%
+r^{2}d\omega ^{2}\text{,}  \label{metd}
\end{equation}%
where $r_{1}$ is an arbitrary constant that affects the normalization of
time. (It is worth noting that, from the formal viewpoint, in the limit $%
r_{0}\rightarrow 0$ our configuration approaches that considered in \cite{z}%
, \cite{bur} but this limit has nothing to do with wormholes.) If we demand,
as usual, $\alpha \leq 1$ ($\alpha =1$ corresponds to the matter that is
stiff with respect to the transversal pressure), it follows from (\ref{ak})
that $\kappa $ lies in the range $1<\kappa \leq 5$.

The metric (\ref{metd}) is not asymptotically flat. However, it can be glued
to the external Schwarzschild solution at some $r_{B}$. In contrast to the
model considered in \cite{sh}, now no restrictions on $r_{B}$ or the
parameter of the model arise since all conditions (\ref{br}) - (\ref{b<})
and (\ref{p0}) - (\ref{p+}) are always satisfied for the solution under
discussion. Matching between (\ref{metd}) and the Schwarzschild region leads
to appearance of the delta-like stresses $T_{\mu }^{\nu (B)}=S_{\mu }^{\nu
}\delta (l-l_{B})$ on the boundary [$l$ is a proper lenght, $l_{B}=l(r_{B})$%
] that plays a role of a singular shell. These stresses are calculated in a
standard way \cite{isr}. If such a shell is massless, the transverse
stresses are equal to $8\pi S_{\theta }^{\theta }=8\pi S_{\phi }^{\phi }=%
\frac{1}{2\sqrt{(1-d)}\sqrt{r_{B}(r_{B}-r_{0})}}>0$. If $r_{B}\rightarrow
\infty $, $S_{\theta }^{\theta }$ becomes negligible.

For $r\gg r_{0}$ we have%
\begin{equation}
ds^{2}=-dt^{2}\frac{r_{1}}{r}+\frac{dr^{2}}{1-d}+r^{2}d\omega ^{2}\text{.}
\label{r1}
\end{equation}

Making a substitution $r=\tilde{r}\sqrt{1-d}$, one can rewrite the metric of
a slice $t=const$ as%
\begin{equation}
dl^{2}=d\tilde{r}^{2}+\tilde{r}^{2}\sqrt{1-d}d\omega ^{2}\text{.}
\end{equation}%
It possesses a deficit of a solid angle in the sense that the area of $%
\tilde{r}=const$ \ is less than $4\pi \tilde{r}^{2}$. The similar phenomenon
occurs, for example, in the gravitational field of a monopole (see page 424
of Ref. \cite{vilbook}). However, it is worth stressing that in our case the
space possesses a wormhole structure with a bridge between two remote
universes. Apart from this, the four-dimensional metric \ref{metd}, \ref{r1}
(that includes also the term with $dt^{2}$) is quite different in that at $%
r=\infty $ a cosmological horizon with an infinite surface area appears on
each side of the bridge, its Hawking temperature being zero. The extremal
horizons with an infinite area were discussed in the black hole context for
the Brance-Dicke theory \cite{broncold}. However, now such a horizon is
cosmological and appears in the framework of general relativity. The
difference consists also in the fact that in \cite{broncold} some
trajectories of particles are able to reach the horizon for a finite proper
time and some are not, whereas in our case the "horizon" is unreachable for
any particle.

To see this, let us consider a particle with an energy $E=-u_{0}$ and
angular momentum $L=u_{\phi }$ (in dimensionless units) moving in the metric %
\ref{met}, (\ref{metd}), $u^{\mu }$ being the four-velocity. Choosing, as
usual, the plane of motion to be $\theta =\frac{\pi }{2}$, one easily
obtains from the condition $g_{\mu \nu }u^{\mu }u^{\nu }=\varepsilon $ $\ $($%
\varepsilon =-1$ for massive particles and $\varepsilon =0$ for massless
ones) that%
\begin{equation}
\left( u^{r}\right) ^{2}=V[\frac{E^{2}}{U}-\frac{L^{2}}{r^{2}}+\varepsilon ]%
\text{,}
\end{equation}%
whence it follows that, irrespective of $\varepsilon $ and particle's
momentum and energy, the proper time needed to reach a horizon, diverges as $%
\tau \sim \sqrt{r}$. Thus, the spacetime is geodesically complete and the
Killing horizon is not an event horizon. It is easy to check that all
curvature components in the orthonormal frame remain finite on this
"horizon".

It is instructive to note that there exists the limit in which $d\rightarrow
1$, $r\rightarrow r_{0}$, $m\rightarrow 0$ in such a way that the proper
distance $l$ from the throat (in this limit $l=\frac{2r_{0}}{\sqrt{1-d}}%
\sqrt{(\frac{r}{r_{0}}-1)}$ in the main approximation) remains finite. Then
the metric turns into%
\begin{equation}
ds^{2}=-dt^{2}+dl^{2}+r_{0}^{2}d\omega ^{2}\text{,}  \label{r0}
\end{equation}%
where we rescaled time according to $t\rightarrow t\sqrt{\frac{r_{1}}{r_{0}}%
\text{.}}$Now all points have $r=const=r_{0}$, $p_{r}=p_{r}^{(0)}=-\rho _{0}$%
, $p_{\perp }=0$, the coordinate $l$ runs in limits ($-\infty $, $\infty $)$%
. $ This is nothing other than string dust source \cite{let}, \cite{dad}.
The effective mass $m_{tot}=\frac{r_{0}}{2}=const$, whereas the proper mass
of matter in the total manifold $m_{p_{r}}=4\pi \int dlr_{0}^{2}\rho $ is
infinite. This is an example of the strong gravitational mass defect that
can be thought of as an analog of similar properties of so-called T-spheres
in the cosmological context \cite{rub}.

To summarize, we have found a simple exact solution of
spherically-symmetrical Einstein equations with a linear (but anisotropic)
equation of state describing a wormhole. The equation of state is linear but
highly anisotropic: $p_{r}<-\rho <0$, $p_{\perp }>0$. At infinity the
spacetime is not asymptotically flat and possesses a regular cosmological
Killing horizon with an infinite area, impenetrable for any particles. This
horizon does not arise if the wormhole region is glued to the Schwarzschild
region. In doing so, the wormhole can enclose an arbitrary amount of the
phantom energy. The configuration under discussion has a limit in which the
phantom energy turns into the string dust, the areal radius tends to the
constant and the strong gravitational mass defect is realized.

\end{document}